\documentclass[12pt]{article}
\usepackage{cite,graphicx,axodraw} % to put in axodraw
\usepackage{latexsym,amssymb,epsf} % don't remember if these are
\newcommand{\newc}{\newcommand}
\newc{\beq}{\begin{equation}}
\newc{\sgn}{\rm{sgn}\,}
\newc{\eeq}{\end{equation}}
\newc{\barr}{\begin{eqnarray}}
\newc{\earr}{\end{eqnarray}}
\newc{\ra}{\rightarrow}
\newc{\lam}{\lambda}
\newc{\eps}{\epsilon}
\newc{\half}{\frac{1}{2}}
\newc{\gev}{\mbox{~GeV}}
\newc{\etal}{{\it et al.}\ }
\newc{\nonum}{\nonumber}
\newc{\kap}{\kappa}
\newc{\rpv}{$/\!\!\!\!R_p $}
\newc{\nonr}{\nonumber}
\newc{\eq}[1]{(\ref{eq:#1})}
\newc{\eqs}[2]{(\ref{eq:#1},\ref{eq:#2})}
\newc{\lab}[1]{\label{eq:#1}}
\newc{\cw}{\cos\theta_w}
\newc{\ssw}{\sin^2\theta_w}
\newc{\upt}{\tilde{u}}
\newc{\elt}{\tilde{\ell}}
\newc{\mut}{\tilde{\mu}}
\newc{\nut}{\tilde{\nu}}
\newc{\dnt}{\tilde{d}}
\newc{\unt}{\tilde{u}}
\newc{\cht}{\tilde{\chi}}
\newc{\psb}{\bar{\psi}}
\newc{\ie}{{\it i.e.}}
\newc{\eg}{{\it e.g.}}

\def\npb#1 #2 #3 #4 {Nucl.~Phys. B {\bf #1}, #2 (#3)#4 }
\def\plb#1 #2 #3 #4 {Phys.~Lett. B {\bf #1}, #2 (#3)#4 }
\def\prd#1 #2 #3 #4 {Phys.~Rev.  D {\bf #1}, #2 (#3)#4 }
\def\prl#1 #2 #3 #4 {Phys.~Rev.~Lett. {\bf #1}, #2 (#3)#4 }
\begin{document}
\title{Attempts at Explaining the NuTeV Observation of Di-Muon Events}
\author{Athanasios Dedes$^1$, Herbi Dreiner$^1$, and 
Peter Richardson$^2$}
\date{\small $^1$ Physikalisches Institut, Universit\"at Bonn, Nu{\ss}allee 
12, D-53115 Bonn, Germany\\
$^2$ DAMTP, Centre for Mathematical Sciences, Wilberforce Road, 
Cambridge CB3 0WA, UK, \\ and 
 Cavendish Laboratory, University of Cambridge, Madingley Road, Cambridge, CB3 0HE, UK} 
\maketitle
%\widetext

\vspace{-9cm}
\begin{flushright}
Cavendish HEP-2001-05\\
DAMTP-2001-47
\end{flushright}
\vspace{8cm}

\begin{abstract}
\noindent
The NuTeV Collaboration has observed an excess in their di-muon channel, 
possibly corresponding to a long-lived neutral particle with only weak 
interactions and which decays to muon pairs. We show that this can {\it not} be 
explained by pair production of neutralinos in the target followed by their 
decay far downstream in the detector via a $LLE$ R-parity violating (RPV) 
operator, as suggested in the literature. In the parameter region allowed by LEP
the event rate is far too small. We propose instead a new neutralino production 
method via $B$-mesons, which can fully explain the observation. This is 
analogous to neutrino production via $\pi$-mesons. This model can be completely 
tested and thus also possibly excluded with NOMAD data. If it is excluded, the 
NuTeV observation is most likely not due to physics beyond the Standard Model. 
Our model can also be tested at the current and future $B$-factories. 
This opens 
up a new way of testing a long-lived neutralino LSP at fixed-target experiments 
and thus a possibility to close the gap between collider and cosmological tests
of R-parity violation. We also discuss a possible explanation in terms of a 
neutral heavy lepton mixing with the Standard Model neutrinos. The flavour 
structure of the observation can be accounted for but the production rate is far 
too low.
\end{abstract}

\vspace*{5mm}
\section{Introduction}
In supersymmetry \cite{Martin:1997ns} with broken R-parity
\cite{Dreiner:1997uz,Bhattacharyya:1997nj} the MSSM superpotential is
extended by
\begin{equation}
W_{\not R_p}=\lam_{ijk}\epsilon_{ab} L_i^aL_j^b{\overline E}_k
+\lam_{ijk}'\epsilon_{ab} L_i^aQ_j^b{\overline D}_k+\lam''_{ijk}
\epsilon_{\alpha\beta\gamma}{\overline U}_i^\alpha{\overline D}_
j^\beta{\overline D}_k^\gamma+\kappa_i\epsilon_{ab}L_i^aH_u^b.
\end{equation}
Here $L,Q$ (${\overline E},\,{\overline U},\,{\overline D}$) are the
lepton and quark doublet (singlet) left-handed chiral superfields,
respectively. $\lam,\lam',\lam''$ are dimensionless coupling
constants, $i,j,k=1,2,3$ are generation indices. $a,b=1,2$ and
$\alpha,\beta,\gamma=1,2,3$ are $SU(2 )_L$ and $SU(3)_c$ gauge
indices, respectively. The main phenomenological changes to the MSSM
are that the lightest supersymmetric particle (LSP) is no longer
stable and supersymmetric particles can be produced singly at
colliders.  Through resonance production the couplings $(\lam,\lam',
\lam'')$ can be probed down to about $10^{-3}$ before the production
cross section becomes too small
\cite{Barger:1989rk,Butterworth:1993tc,Dreiner:2001vf,Kalinowski:1997bc,Erler:1997ww}.
If we consider MSSM supersymmetric pair production with a neutralino
LSP then we can typically probe couplings down to $10^{-5}$ or $10^{
-6}$ \cite{Dreiner:1991pe,Roy:1992qr,Dreiner:1997cd,Allanach:2001xz}.
For smaller couplings the LSP decays outside the detector and we
retrieve the MSSM signatures at colliders. Cosmologically one can
exclude lifetimes for the LSP between
$1\,\rm{s}<\tau_{\chi^0_1}<10^{17}\,\rm{yr}$ \cite{Ellis:1992nb},
which corresponds to couplings $10^{-22}<(\lam,
\lam',\lam'')<10^{-10}$.  This leaves a gap in experimental
sensitivity to the R-parity violating couplings\footnote{These
coupling values have been determined for a photino LSP of $M_{\chi^0
_1}={\cal O}(50) \gev$ and scalar fermion masses of $M_{\tilde
f}={\cal O}(100 \gev)$.}  $10^{-10}<(\lam,\lam',\lam'')<10^{-6}$
\cite{Dreiner:1997uz}. Fixed-target experiments with remote detectors
can probe significantly longer lifetimes than collider experiments and
are thus an ideal environment for closing this gap in sensitivity
\cite{Borissov:2000eu}.

The NuTeV Collaboration has searched for long-lived neutral particles
($N^0$) with mass $M_{N^0}\geq2.2\gev$ and small interaction rates
with ordinary matter
\cite{Formaggio:2000ne,Adams:2000rd,Adams:2001sk}.  They look for the
decay of the neutral particles in a detector which is $1.4\,\rm{km}$
downstream from the production point. They observe 3 $\mu\mu$ events
where they only expect to see a background of $0.069\pm0.010$ events.
The probability that this is a fluctuation of this specific channel is
about $8\cdot10^{-5}$, which corresponds to about $4.6\,\sigma$. The
probability for a fluctuation of this magnitude into any of the
di-lepton channels is about $3\,\sigma$.

The NuTeV experiment considered in detail the possibility that this
discrepancy is due to a $N^0$ which decays into a three-body final
state. In Ref.~\cite{Adams:2000rd} several kinematic distributions of
the di-muon events were checked against the hypothesis of a $N^0$ with
mass $5\,\rm{GeV}$: the transverse mass-, invariant di-muon mass- and
the missing $p_T$-distributions all agree well with the $N^0$
hypothesis.  The distribution in the energy asymmetry \mbox{$A_E\equiv
  |E_1 -E_2|/(E_1+E_2)$} of the three events ($E_1$ and $E_2$ are the
two observed muon energies in each event) shows a low probability for
the $N^0$ hypothesis. Thus three out of four distributions work very
well and as NuTeV does, we consider it worthwhile to investigate
whether this observation could be due to new physics.  It is the
purpose of this letter to consider two possible models which could
explain the observation: {\it (i)} a light neutralino which decays via
R-parity violation and {\it (ii)} a neutral heavy lepton (NHL) mixing
with the Standard Model neutrinos.

A search for R-parity violating neutralino decays at NuTeV has been
proposed in Ref.~\cite{Borissov:2000eu} and the couplings $\lam_{122}$
and $\lam_{133}$ were discussed. In
Ref.~\cite{Adams:2000rd,Adams:2001sk} NuTeV themselves mention the
possibility of R-parity violating neutralino decays as a solution to
the observed discrepancy, without looking at any specific couplings.
In Ref.~\cite{Formaggio:2000ne} NuTeV searched for the neutralino of a
very specific model \cite{Choudhury:2000tn}. This neutralino was very
light and decayed via $L_1L_2{\bar E}_1$ or $L_1L_3{\bar E}_1$ to an
$ee$ final state.  Certain supersymmetric parameter ranges were
excluded assuming neutralino pair-production.

Here we show that the simple scenarios discussed in the literature can
not lead to an excess at NuTeV, since the decisive supersymmetric
parameter range to get a significant neutralino production cross
section has been excluded by LEP. We propose instead the production of
light neutralinos via $B$-mesons which could give a measurable excess.
We briefly present the two possible models and then discuss them
quantitatively.

In section \ref{sec:nhl} we show that the production rate for neutral
heavy leptons is also too low and does not lead to a viable
explanation.

%%%%%%%%%%DIAGRAMS HERE%%%%%%%%%%%%%%%%%%%%%
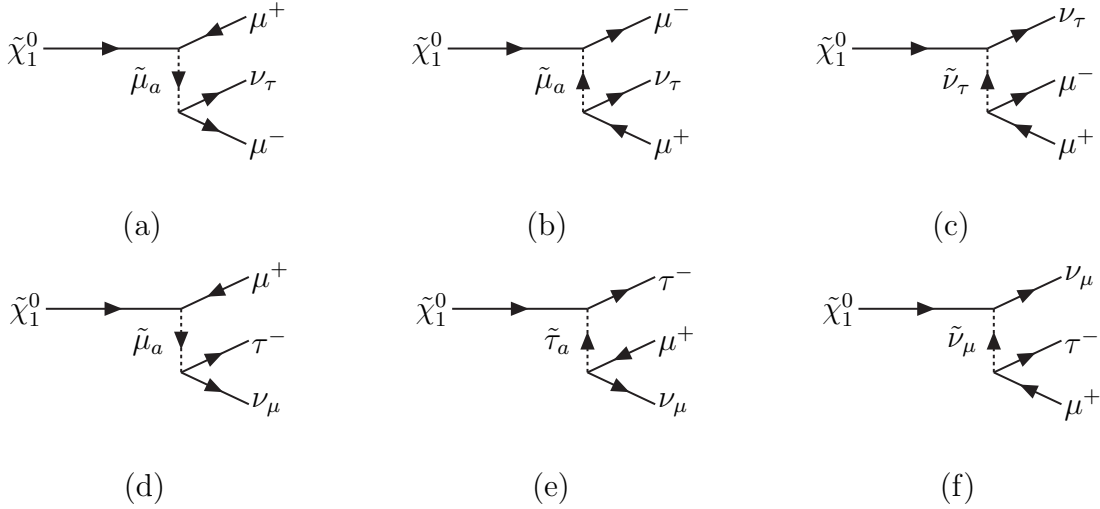
\begin{figure}
\begin{center}
%%%%%%%%%%%%%%%%%%%%%%%%%%%%%%%%%%%%%%%%
% diagrams for process chi_1^0 -> muon,anti-muon,tau-neutrino            
%%%%%%%%%%%%%%%%%%%%%%%%%%%%%%%%%%%%%%%
{
\unitlength=1.5 pt
\SetScale{1.5}
\SetWidth{0.5}      % line    size control
%\tiny    %  letter  size control
{} \qquad\allowbreak
%  diagram # 1
\begin{picture}(79,65)(0,0)
\Text(13.0,49.0)[r]{$\tilde{\chi}_1^0$}
\ArrowLine(14.0,49.0)(48.0,49.0) 
\Text(66.0,57.0)[l]{${\mu}^+$}
\ArrowLine(65.0,57.0)(48.0,49.0) 
\Text(44.0,41.0)[r]{$\tilde{\mu}_a$}
\DashArrowLine(48.0,49.0)(48.0,33.0){1.0} 
\Text(66.0,41.0)[l]{$\nu_\tau$}
\ArrowLine(48.0,33.0)(65.0,41.0) 
\Text(66.0,25.0)[l]{$\mu^-$}
\ArrowLine(48.0,33.0)(65.0,25.0) 
\Text(39,0)[b] {(a)}
\end{picture} \ 
{} \qquad\allowbreak
%  diagram # 2
\begin{picture}(79,65)(0,0)
\Text(13.0,49.0)[r]{$\tilde{\chi}_1^0$}
\ArrowLine(14.0,49.0)(48.0,49.0) 
\Text(66.0,57.0)[l]{$\mu^-$}
\ArrowLine(48.0,49.0)(65.0,57.0) 
\Text(44.0,41.0)[r]{$\tilde{\mu}_a$}
\DashArrowLine(48.0,33.0)(48.0,49.0){1.0} 
\Text(66.0,41.0)[l]{$\nu_\tau$}
\ArrowLine(48.0,33.0)(65.0,41.0) 
\Text(66.0,25.0)[l]{${\mu}^+$}
\ArrowLine(65.0,25.0)(48.0,33.0) 
\Text(39,0)[b] {(b)}
\end{picture} \ 
{} \qquad\allowbreak
%  diagram # 3
\begin{picture}(79,65)(0,0)
\Text(13.0,49.0)[r]{$\tilde{\chi}_1^0$}
\ArrowLine(14.0,49.0)(48.0,49.0) 
\Text(66.0,57.0)[l]{$\nu_\tau$}
\ArrowLine(48.0,49.0)(65.0,57.0) 
\Text(44.0,41.0)[r]{$\tilde{\nu}_\tau$}
\DashArrowLine(48.0,33.0)(48.0,49.0){1.0} 
\Text(66.0,41.0)[l]{$\mu^-$}
\ArrowLine(48.0,33.0)(65.0,41.0) 
\Text(66.0,25.0)[l]{${\mu}^+$}
\ArrowLine(65.0,25.0)(48.0,33.0) 
\Text(39,0)[b] {(c)}
\end{picture} \ 
}

%%%%%%%%%%%%%%%%%%%%%%%%%%%%%%%%%%%%%%%
% diagrams for process chi_1^0 -> muon-neutrino,anti-muon,tau 
%%%%%%%%%%%%%%%%%%%%%%%%%%%%%%%%%%%%%%           
{
\unitlength=1.5 pt
\SetScale{1.5}
\SetWidth{0.5}      % line    size control
%\tiny    %  letter  size control
%{} \qquad\allowbreak
%  diagram # 1
{} \qquad\allowbreak
\begin{picture}(79,65)(0,0)
\Text(13.0,49.0)[r]{$\tilde{\chi}_1^0$}
\ArrowLine(14.0,49.0)(48.0,49.0) 
\Text(66.0,57.0)[l]{${\mu}^+$}
\ArrowLine(65.0,57.0)(48.0,49.0) 
\Text(44.0,41.0)[r]{$\tilde{\mu}_a$}
\DashArrowLine(48.0,49.0)(48.0,33.0){1.0} 
\Text(66.0,41.0)[l]{$\tau^-$}
\ArrowLine(48.0,33.0)(65.0,41.0) 
\Text(66.0,25.0)[l]{$\nu_\mu$}
\ArrowLine(48.0,33.0)(65.0,25.0) 
\Text(39,0)[b] {(d)}
\end{picture} \
%%%%%%%%%%%%%%%%%%%%%%%%%%%%%%%5
{} \qquad\allowbreak
%  diagram # 2
\begin{picture}(79,65)(0,0)
\Text(13.0,49.0)[r]{$\tilde{\chi}_1^0$}
\ArrowLine(14.0,49.0)(48.0,49.0) 
\Text(66.0,57.0)[l]{$\tau^-$}
\ArrowLine(48.0,49.0)(65.0,57.0) 
\Text(44.0,41.0)[r]{$\tilde{\tau}_a$}
\DashArrowLine(48.0,33.0)(48.0,49.0){1.0} 
\Text(66.0,41.0)[l]{${\mu}^+$}
\ArrowLine(65.0,41.0)(48.0,33.0) 
\Text(66.0,25.0)[l]{$\nu_\mu$}
\ArrowLine(48.0,33.0)(65.0,25.0) 
\Text(39,0)[b] {(e)}
\end{picture} \ 
{} \qquad\allowbreak
%  diagram # 3
\begin{picture}(79,65)(0,0)
\Text(13.0,49.0)[r]{$\tilde{\chi}_1^0$}
\ArrowLine(14.0,49.0)(48.0,49.0) 
\Text(66.0,57.0)[l]{$\nu_\mu$}
\ArrowLine(48.0,49.0)(65.0,57.0) 
\Text(44.0,41.0)[r]{$\tilde{\nu}_\mu$}
\DashArrowLine(48.0,33.0)(48.0,49.0){1.0} 
\Text(66.0,41.0)[l]{$\tau^-$}
\ArrowLine(48.0,33.0)(65.0,41.0) 
\Text(66.0,25.0)[l]{${\mu}^+$}
\ArrowLine(65.0,25.0)(48.0,33.0) 
\Text(39,0)[b] {(f)}
\end{picture} \ 
}
\end{center}
%\end{document}
\caption{Neutralino decays through the R-parity violating coupling 
$\lambda_{232}$.
Diagrams (a-c) give rise to di-muon events  while
the diagrams (d-f) to tau-muon ones. The index a=1,2 denotes the mass
eigenstate of the slepton.}
\end{figure}
%%%%%%%%%%%%%%%%%%%%%%%%%%%%%%%%%%%%%%%%%%%%%
%%%%%%%%%%DIAGRAMS END HERE%%%%%%%%%%%%%%%%%%

\section{The $R_p$ Violating Model}
\label{sec:model}
The heavy neutral particle we consider is the lightest neutralino
$\chi^0_1$, which we also assume to be the LSP. In the notation of
\cite{Dreiner:1991pe}, the neutralino decays as $\chi^0_1\ra{\cal
  O}_{\not \!R_p}$, where ${\cal O}_{\not \!R_p}$ is the dominant
R-parity violating operator. Only two operators give a di-muon
signature: $\lam_{2i2}\epsilon_{ab}L^a_2 L^b_i{\bar E}_2$, $i=1,3$.
For $i=1$ the neutralino will decay with equal probability to
$e\mu\nu$ and $\mu\mu\nu$. No $e\mu$-events are observed, we therefore
propose one dominant R-parity violating operator:
\begin{equation}
{\cal O}_{\not \!R_p}=\epsilon_{ab} \lam_{232} L_\mu^a L_\tau^b {\bar E}_\mu.
\label{eq:operator}
\end{equation}
For later reference we quote the experimental bound on this operator
\cite{Allanach:1999ic}
\begin{equation}
 \lam_{232}< 0.070
\left(\frac{m_{{\tilde \mu}_R}}{100\,\rm{GeV}}\right),\quad (2\,\sigma).
\end{equation}
The operator in Eq.(\ref{eq:operator}) corresponds 
to the two neutralino decay modes (Fig.1)
\begin{eqnarray}
\chi^0_1&\rightarrow& \left\{ \begin{array}{c}
\mu^-_L\mu^+_R\nu_\tau, \\
\tau^-_L\mu^+_R\nu_\mu, \\
\end{array}\right.
\label{eq:decays}
\end{eqnarray}
as well as their complex conjugate, since the neutralino is a Majorana
spinor. We shall show below that for a light neutralino the $\tau\mu$ decays
are sufficiently phase space suppressed to give an expectation below
one event. For the light neutralino production we shall consider two
possibilities:
\begin{enumerate}
\item Pair production of the neutralinos \cite{Bartl:1986hp} which
  proceeds via (a) s-channel $Z^0$ boson exchange and (b) t-channel
  squark exchange.
\item Single neutralino production in the decay of bottom hadrons. The
bottom hadrons are formed following the production of a $b\bar{b}$
pair. These hadrons can then decay via the R-parity couplings
$\lambda'_{i13}$, ($i=1,2,3$). We will only consider the decays of the
$B^0_{d}$ and $B^+$ via R-parity violation (Fig.2)
\begin{eqnarray}
B^0_d&\longrightarrow& \bar{\nu}_i \cht^0_1,\\
B^+ &\longrightarrow &\ell^+_i \cht^0_1.
\end{eqnarray}
This mechanism allows one to produce light neutralinos via a strong
interaction process and is analogous to the production of neutrino
beams via $\pi$'s and $K$'s (and $D$'s). A related mechanism was
discussed in the context of the Karmen time anomaly
\cite{Choudhury:1996pj,Choudhury:2000tn}.
\end{enumerate}
For later reference we present the experimental bounds on the $\lam'_
{i13}$ at $2\,\sigma$ \cite{Allanach:1999ic,Bhattacharyya:1997nj}
\begin{equation}
\lam'_{113}< 0.021 \frac{m_{{\tilde b}_R}}{100\,\rm{GeV}},\quad
\lam'_{213}< 0.059 \frac{m_{{\tilde b}_R}}{100\,\rm{GeV}},\quad
\lam'_{313}< 0.11 \frac{m_{{\tilde b}_R}}{100\,\rm{GeV}}.\label{eq:bounds}
\end{equation}

%%%%%%%%%%%DIAGRAMS HERE%%%%%%%%%%%%%%%%%%%
% diagrams for process d,B -> n2,~o1              
\begin{figure}
\begin{center}
{
\unitlength=1.5 pt
\SetScale{1.5}
\SetWidth{0.5}      % line    size control
%\tiny    %  letter  size control
{} \qquad\allowbreak
%  diagram # 1
\begin{picture}(79,65)(0,0)
\Text(13.0,57.0)[r]{$d$}
\ArrowLine(14.0,57.0)(31.0,49.0) 
\Text(13.0,41.0)[r]{$\bar{b}$}
\ArrowLine(31.0,49.0)(14.0,41.0) 
\Text(39.0,53.0)[b]{$\tilde{\nu}^*_i$}
\DashArrowLine(48.0,49.0)(31.0,49.0){1.0} 
\Text(66.0,57.0)[l]{$\tilde{\chi}_1^0$}
\Line(48.0,49.0)(65.0,57.0) 
\Text(66.0,41.0)[l]{$\bar{\nu}_i$}
\ArrowLine(65.0,41.0)(48.0,49.0) 
\Text(39,20)[b] {(a)}
\end{picture} \ 
{} \qquad\allowbreak
%  diagram # 2
\begin{picture}(79,65)(0,0)
\Text(13.0,57.0)[r]{$d$}
\ArrowLine(14.0,57.0)(48.0,57.0) 
\Text(66.0,57.0)[l]{$\tilde{\chi}_1^0$}
\Line(48.0,57.0)(65.0,57.0) 
\Text(44.0,49.0)[r]{$\tilde{d}_L$}
\DashArrowLine(48.0,57.0)(48.0,41.0){1.0} 
\Text(13.0,41.0)[r]{$\bar{b}$}
\ArrowLine(48.0,41.0)(14.0,41.0) 
\Text(66.0,41.0)[l]{$\bar{\nu}_i$}
\ArrowLine(65.0,41.0)(48.0,41.0)
\Text(39,20)[b] {(b)}
\end{picture} \ 
{} \qquad\allowbreak
%  diagram # 3
\begin{picture}(79,65)(0,0)
\Text(13.0,57.0)[r]{$d$}
\ArrowLine(14.0,57.0)(48.0,57.0) 
\Text(66.0,57.0)[l]{$\bar{\nu}_i$}
\ArrowLine(65.0,57.0)(48.0,57.0)
\Text(44.0,49.0)[r]{$\tilde{b}_R$}
\DashArrowLine(48.0,57.0)(48.0,41.0){1.0} 
\Text(13.0,41.0)[r]{$\bar{b}$}
\ArrowLine(48.0,41.0)(14.0,41.0) 
\Text(66.0,41.0)[l]{$\tilde{\chi}_1^0$}
\Line(48.0,41.0)(65.0,41.0) 
\Text(39,20)[b] {(c)}
\end{picture} \ 

{} \qquad\allowbreak %\qquad\allowbreak
\begin{picture}(79,65)(0,0)
\Text(13.0,57.0)[r]{$u$}
\ArrowLine(14.0,57.0)(31.0,49.0) 
\Text(13.0,41.0)[r]{$\bar{b}$}
\ArrowLine(31.0,49.0)(14.0,41.0) 
\Text(39.0,53.0)[b]{$\tilde{\ell}_i$}
\DashArrowLine(48.0,49.0)(31.0,49.0){1.0} 
\Text(66.0,57.0)[l]{$\tilde{\chi}_1^0$}
\Line(48.0,49.0)(65.0,57.0) 
\Text(66.0,41.0)[l]{${\ell}^+_i$}
\ArrowLine(65.0,41.0)(48.0,49.0) 
\Text(39,20)[b] {(d)}
\end{picture} \ 
{} \qquad\allowbreak
%  diagram # 3
\begin{picture}(79,65)(0,0)
\Text(13.0,57.0)[r]{$u$}
\ArrowLine(14.0,57.0)(48.0,57.0) 
\Text(66.0,57.0)[l]{$\tilde{\chi}_1^0$}
\Line(48.0,57.0)(65.0,57.0) 
\Text(44.0,49.0)[r]{$\tilde{u}_L$}
\DashArrowLine(48.0,57.0)(48.0,41.0){1.0} 
\Text(13.0,41.0)[r]{$\bar{b}$}
\ArrowLine(48.0,41.0)(14.0,41.0) 
\Text(66.0,41.0)[l]{${\ell}_i^+$}
\ArrowLine(65.0,41.0)(48.0,41.0) 
\Text(39,20)[b] {(e)}
\end{picture} \ 
{} \qquad\allowbreak
%  diagram # 4
\begin{picture}(79,65)(0,0)
\Text(13.0,57.0)[r]{$u$}
\ArrowLine(14.0,57.0)(48.0,57.0) 
\Text(66.0,57.0)[l]{${\ell}_i^+$}
\ArrowLine(65.0,57.0)(48.0,57.0) 
\Text(44.0,49.0)[r]{$\tilde{b}_R$}
\DashArrowLine(48.0,57.0)(48.0,41.0){1.0} 
\Text(13.0,41.0)[r]{$\bar{b}$}
\ArrowLine(48.0,41.0)(14.0,41.0) 
\Text(66.0,41.0)[l]{$\tilde{\chi}_1^0$}
\Line(48.0,41.0)(65.0,41.0) 
\Text(39,20)[b] {(f)}
\end{picture} \ 
%{} \qquad\allowbreak 
}
\vspace{-18mm}
\end{center}
\caption{Neutralino production in $B$-meson decays : (a-c)
$B^0_d \longrightarrow \bar{\nu}_i \cht^0_1$, and (d-f)
$B^+ \longrightarrow \ell^+_i \cht^0_1$.}
\end{figure}
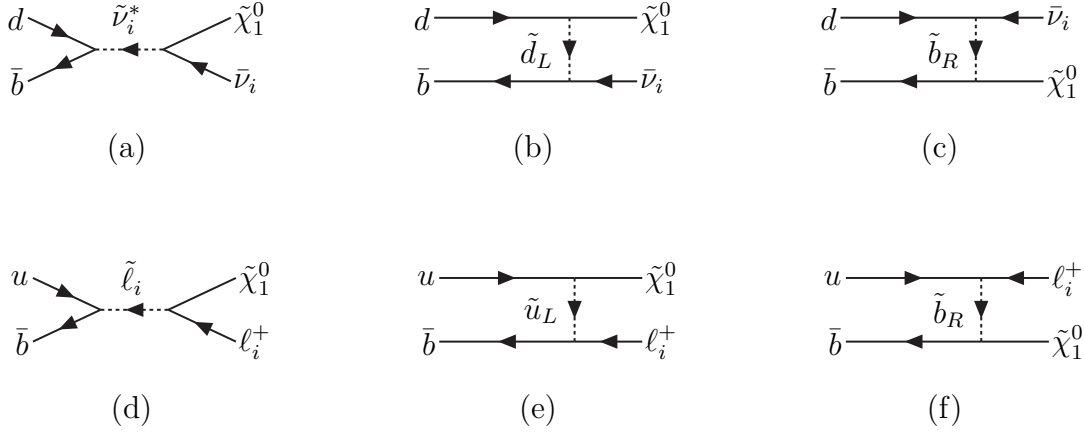
%%%%%%%%%%%%%%%END DIAGRAMS%%%%%%%%%%%%%%%%%%%%%%

\section{Quantitative Analysis}

\begin{figure}[t]
\begin{center}
\includegraphics[width=0.7\textwidth]{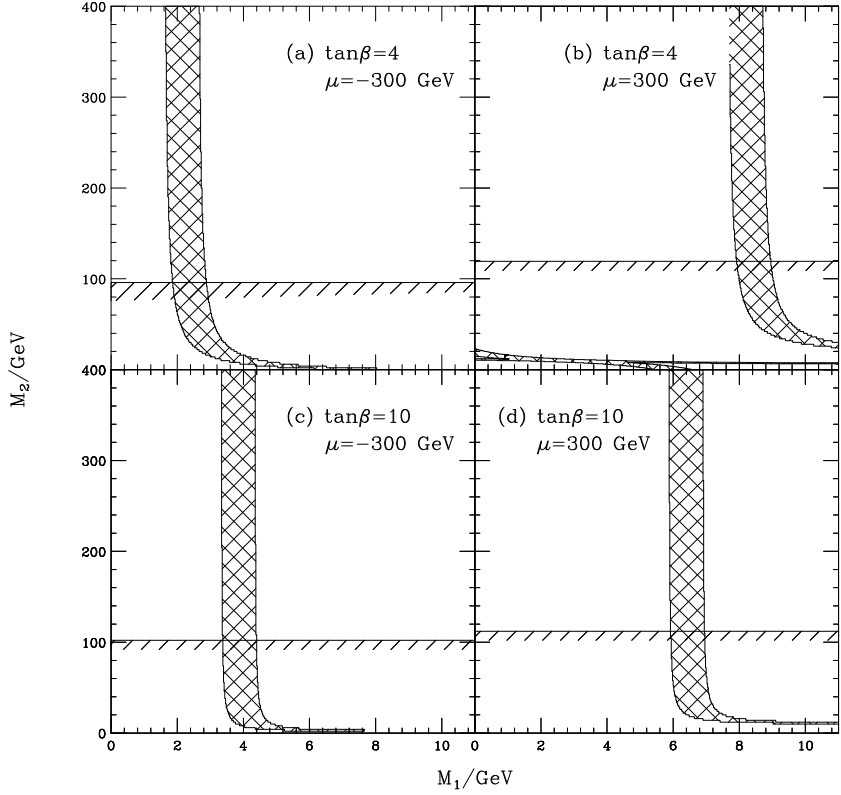}
\end{center}
\caption{Solutions in ($M_1$, $M_2$, $\mu$, $\tan \beta$) giving 
        $4.5\,\rm{GeV}\leq M_{\cht^0_1}\leq 5.5\, \rm{GeV}$ in the
         cross-hatched region. Points below the horizontal hatched line are 
        excluded by the requirement that $M_{\cht^+_1}>100\,\rm{GeV}$.}
\label{fig:mssmparam}
\end{figure}

As discussed by the NuTeV experiment, the mass of the $N^0$ is roughly
$5\gev$. The constraints on a very light neutralino were discussed in
detail in Ref.\cite{Choudhury:2000tn}. We expect them to mainly carry
over to the present mass region \cite{peterR}. In order to get a
\mbox{$M_{\chi^0_1}={\cal O}(5\gev)$} neutralino and avoid the LEP
bounds we must consider the case, where the electroweak gaugino masses
$M_1,\,M_2$ are independent parameters. In Fig.\,\ref{fig:mssmparam}
we show the MSSM parameter space which corresponds to $M_{\chi^0_1}
=(5\pm0.5)\gev$ for two values of $\tan\beta$ and $\sgn\mu$. The
composition of the neutralino is more than $99\%$ bino, provided the
lightest chargino mass is greater than $100\,\rm{GeV}$.

The dominant bino-nature of the LSP has immediate implications for
pair production of neutralinos. The bino does not couple to the $Z^0$
boson and thus the s-channel pair-production of the bino is
negligible. This only leaves the t-channel production which is
proportional to $M_{{\tilde q}}^{-4}$ and thus strongly suppressed.
We shall quantify this below.

In both cases neutralino production is followed by the decay.  The
matrix elements for the decay via~~\rpv\ were given in
\cite{Dreiner:2000qz,Dreiner:1997cd}. As the neutralino in our model
will be much lighter than the sleptons ($M_{{\tilde\ell}}\gtrsim
90\,\rm{GeV}$ from LEP) it is sufficient to neglect the momentum flow
through the slepton propagators. For a purely bino neutralino in this
limit the spin averaged matrix element is given by
%%%%%%%%%%%%%%%%%%%%%%%%%%%%%%%%%%%%%%%%%%%
\begin{eqnarray}
\lefteqn{|\overline{\mathcal{M}}|^2(\cht^0_1\ra{\bar{\nu_i}\ell^+_j\ell^-_k})=}&&\\
&& \frac{g'^2\lambda^2_{ijk}}{4}\left[\rule{0mm}{7mm}
  \frac{Y^2_{\nu_i}}{M^4_{\nut_i}}
  \left(m^2_{\ell_j\ell_k}-m^2_{\ell_j}-m^2_{\ell_k}\right)
  \left(M^2_{\cht^0_1}-m^2_{\ell_j\ell_k}\right)
  -2\frac{Y_{\nu_i}Y_{\ell_{jL}}}{M^2_{\nut_i}M^2_{\elt_{jL}}}
  \left(m^2_{\nu_i\ell_k}m^2_{\ell_j\ell_k}-M^2_{\cht^0_1m^2_{\ell_k}}\right)
\right.\nonumber\\
&& +\frac{Y^2_{\ell_{jL}}}{M^4_{\elt_{jL}}}
\left(m^2_{\nu_i\ell_k}-m^2_{\ell_k}\right)
\left(M^2_{\cht^0_1}+m^2_{\ell_j}-m^2_{\nu_i\ell_k}\right)
+2\frac{Y_{\nu_i}Y_{\ell_{kR}}}{M^2_{\nut_i}M^2_{\elt_{kR}}}
\left(m^2_{\nu_i\ell_j}m^2_{\ell_j\ell_k}-M^2_{\cht^0_1m^2_{\ell_j}}\right)
\nonumber\\
&& \left.  +\frac{Y^2_{\ell_{kR}}}{M^4_{\elt_{kR}}}
  \left(m^2_{\nu_i\ell_j}-m^2_{\ell_j}\right)
  \left(M^2_{\cht^0_1}+m^2_{\ell_k}-m^2_{\nu_i\ell_j}\right)
  +2\frac{Y_{\ell_{jL}}Y_{\ell_{kR}}}{M^2_{\elt_{jL}}M^2_{\elt_{kR}}}
  \left(m^2_{\nu_i\ell_j}m^2_{\nu_i\ell_k}-m^2_{\ell_j}m^2_{\ell_k}\right)
  \rule{0mm}{7mm}\right]\!.\nonumber 
\end{eqnarray} 
Here $Y_f$ is the hypercharge of the field $f$ and $m_{f_if_j}=
(f_i+f_j) ^2$ is the invariant mass of the $f_i$, $f_j$ pair of
fields. This matrix element can be simplified by assuming a common
sfermion mass, $M_{\tilde{f}}$, and by putting in explicit values for
the couplings
\beq
|\overline{\mathcal{M}}|^2(\cht^0_1\ra\bar{\nu_i}\ell^+_j\ell^-_k)=
\frac{9g'^2\lambda^2_{ijk}}{4M^4_{\tilde{f}}}
\left(M^2_{\cht^0_1}+m^2_{\ell_k}-m^2_{\nu_i\ell_j}\right)
\left(m^2_{\nu_i\ell_j}-m^2_{\ell_j}\right).  
\label{eq:neutdecay}
\eeq

In the analysis of Ref.~\cite{Adams:2000rd,Adams:2001sk} the model for
the heavy neutral lepton decay studied was based on a weak decay
matrix element \cite{Formaggio:1998zn} 
\beq
\mathcal{M}(N^0\ra\nu_i\ell^-_j\ell^+_k) = \frac{G_F}{\sqrt{2}}
\bar{u}_{N^0} \gamma^\mu(1-\gamma_5)u_{\ell_j}
\bar{v}_{\ell_i}\gamma_\mu(1-\gamma_5)u_{\nu_i}.  
\end{equation}
If we compute the squared amplitude and average over the spin of the
incoming heavy lepton we obtain 
\begin{equation}
|\overline{\mathcal{M}}|^2(N^0\ra\nu_i\ell^-_j\ell^+_k) =
16G^2_F\left(m^2_{N^0}+m^2_{\ell_k}-m^2_{\nu_i\ell_j}\right)
\left(m^2_{\nu_i\ell_j}-m^2_{\ell_j}\right).  
\end{equation} 
So the distribution of the decay products from the R-parity violating
decay will be exactly the same as the weak decay matrix element
studied in \cite{Adams:2000rd,Adams:2001sk} and therefore this model
has exactly the same problem with the energy asymmetry $A_E$ as that
discussed in \cite{Adams:2000rd,Adams:2001sk}.

\subsection{Neutralino Pair Production}

\begin{figure}
\centerline{\includegraphics[width=0.65\textwidth,angle=90]{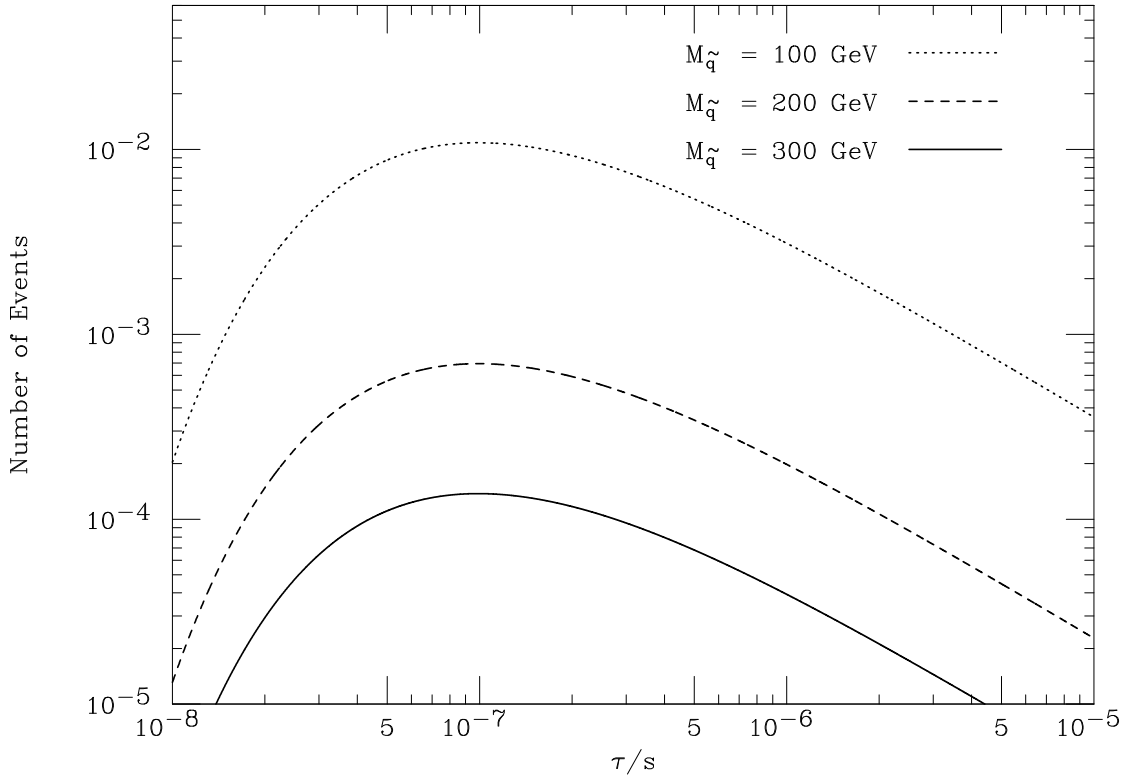}}
\caption{Number of Events in the NuTeV detector for neutralino pair production
         as a function of the neutralino lifetime.}
\label{fig:pair}
\end{figure}

We simulated neutralino pair production using
HERWIG~6.2\cite{Corcella:2001bw,Corcella:1999qn,Marchesini:1992ch}.\footnote{
  One modification to HERWIG was made in that we used the average of
  the central and higher gluon parton distribution functions from the
  leading-order fit of \cite{Martin:1998np}. This will become the
  default in the next release of HERWIG.}  This allows us to simulate
the production cross section with the correct momentum spectrum for
the neutralinos and to determine whether they can decay within the
NuTeV detector.  Those events where the neutralino could decay in the
detector were weighted with the probability that the neutralino
decayed in the detector, for a given lifetime 
\begin{equation}
{\cal P}\approx exp\left\{{-\frac{\ell}{\beta\gamma c\tau_{\chi^0}}}
\right\}
\frac{\Delta x}{\beta\gamma c\tau_{\chi^0}},
\label{eq:decayprob}
\end{equation}
where $\ell=1.4\,\rm{km}$ is the distance target-detector, $\Delta
x=35\,\rm{m}$ is the length of the detector, $\beta c$ is the speed of
the neutralino and $\tau_{\chi^0}$ is its lifetime. The neutralino was
furthermore decayed with the full RPV matrix element
\cite{Richardson:2000nt,Dreiner:2000qz}. We then applied the NuTeV
kinematic cuts \cite{Adams:2001sk} on the neutralino decay products.
We required that the neutralinos decay within the fiducial
volume\footnote{In the original version of our paper this number was
smaller as found in \cite{preprint}. We thank T.~Adams for drawing our
attention to the corrected value in the published version
\cite{Adams:2001sk}.} \mbox{$(2.54\times2.54\times28)$~m$^3$} of the
NuTeV detector at a distance $1.4\,\rm{km}$ downstream of the
production target. The muons produced in the neutralino decay were
required to have energy, $E_\mu >2.2\,\rm{GeV}$, and the transverse
mass, \mbox{$m_T=|P_T|+\sqrt{P^2_T+M^2_V}>2.2\gev$}, as in
Ref.\cite{Formaggio:2000ne,Adams:2000rd, Adams:2001sk}. Here $P_T$ and
$M_T$ are the transverse momentum and mass of the visible decay
products, respectively.
  
As the production of a bino only occurs via $t$-channel squark
exchange the cross section will depend on the (assumed degenerate)
squark mass as $\sim1/M^4_{\tilde{q}}$. The number of events which
would be observed in the NuTeV detector are given in
Fig.\,\ref{fig:pair} as a function of the lifetime of the neutralino.
Given the current limits on the squark mass from both LEP
\cite{Barate:2001tu,Acciarri:1999xb,Abreu:2000qj,Abbiendi:1999yz} and
the Tevatron \cite{Abe:1997yy,Abbott:1999xc} it is impossible, for any
neutralino lifetime, to get sufficient events to explain the NuTeV
results via neutralino pair production. In Ref.\cite{Borissov:2000eu}
the LEP constraints on the MSSM parameter space were not taken into
account.

\subsection{Neutralino Production in $B$-meson Decays}

\begin{figure}[t!]
\begin{center}
\includegraphics[width=0.65\textwidth,angle=90]{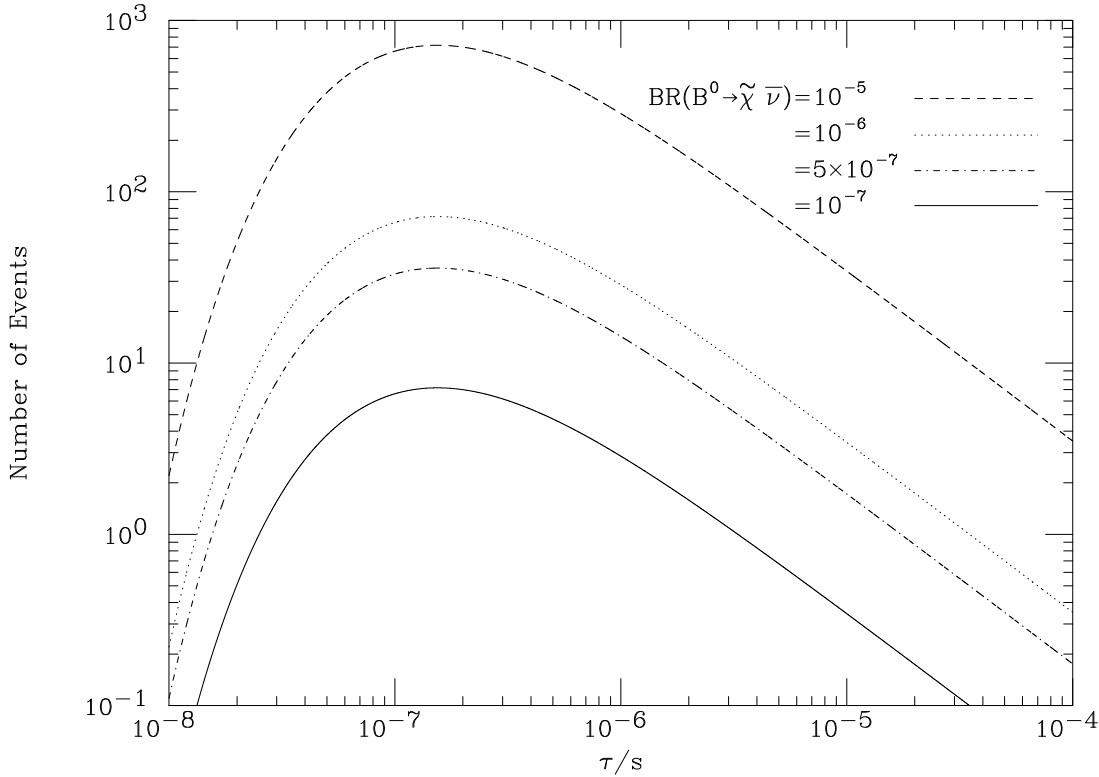}
\caption{Number of events in the NuTeV detector for neutralino production in 
  $B$-meson decays as a function of the neutralino lifetime.}
\label{fig:Bdecay}
\end{center}
\end{figure}
\begin{figure}[t]
\begin{center}
  \includegraphics[width=0.7\textwidth,angle=90]{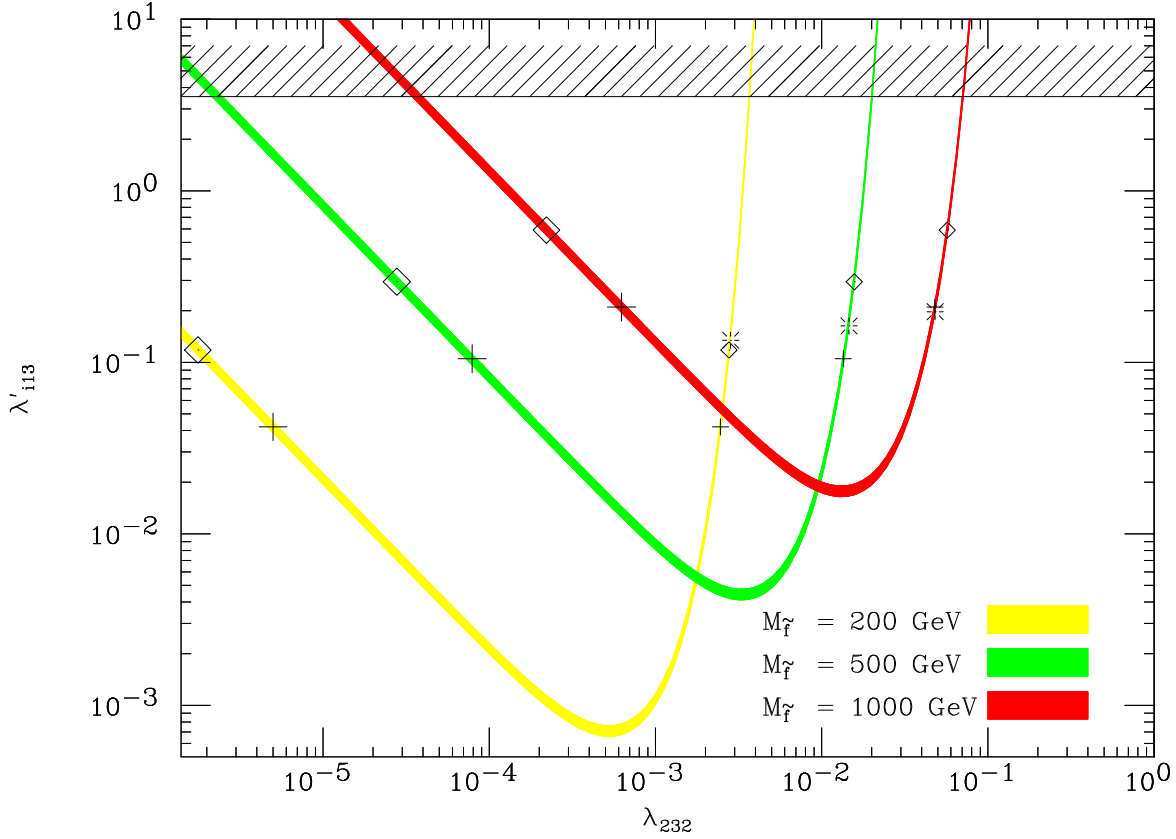}
\caption{Regions in $\lambda_{232},\lambda'_{i13}$ parameter space in which
  we would expect $3\pm1$ events to be observed in the NuTeV detector.
  The limits from \cite{Allanach:1999ic,Bhattacharyya:1997nj} on the
couplings $\lam'_{113}$ (crosses) and $\lam'_{213}$ (diamonds) allow
solutions between the two points for each of the masses shown.  The
region above the stars is ruled out for the coupling $\lam'_{213}$ by
the limit on the product of the couplings $\lam_{232}\lam'_{213}$ from
the limit on the branching ratio $B^+\ra\mu^+\nu$ \cite{Groom:2000in}.
The hatched region shows the experimental bound on the coupling
$\lam'_{i13}$ from perturbativity.  The corresponding limits on the
coupling $\lam_{232}$ from both low energy experiments
\cite{Allanach:1999ic,Bhattacharyya:1997nj} and perturbativity are
not shown as our solutions do not extend into this region.}
\label{fig:parameter}
\end{center}
\end{figure}

As with the neutralino pair production we used HERWIG to simulate
$b\bar{b}$ production. One of the produced $B$ mesons was then forced
at random to decay via RPV. The overall normalization was properly
taken into account.  The partial widths for the decays of the $B_0$
and $B^+$ via RPV are given by 
\begin{eqnarray}
\Gamma\left(B^0_{d}\ra\bar{\nu}_i\cht^0_1\right) &=&
\frac{\lam'^2_{i13}f^2_Bm^2_{B^0}p_{cm}}{16\pi(m_d+m_b)^2}
\left[\frac{L_{\nu_i}}{M^2_{\tilde{\nu}_i}}
  -\frac{L_d}{2M^2_{\dnt_L}}-\frac{R_b^*}{2M^2_{\tilde{b}_R}}\right]^2
\left(M^2_{B^0}-M^2_{\cht^0_1}\right), \nonumber\\
&=&\frac{9\lam'^2_{i13}g'^2f^2_Bm^2_{B^0}p_{cm}}{256\pi(m_d+m_b)^2M^4_{\tilde{f}}}
\left(M^2_{B^0}-M^2_{\cht^0_1}\right),  \label{eqn:B0}\\
\Gamma\left(B^+\ra\ell^+_i\cht^0_i\right) &=&
\frac{\lam'^2_{i13}f^2_Bm^2_{B^+}p_{cm}}{8\pi(m_u+m_b)^2}
\left[\frac{L_{\ell_i}}{M^2_{\elt_i}}
  -\frac{L_u}{2M^2_{\upt_L}}-\frac{R_b^*}{2M^2_{\tilde{b}_R}}\right]^2
\left(M^2_{B^+}-m^2_{\ell_i}-M^2_{\cht^0_1}\right),\nonumber\\
&=&\frac{9\lam'^2_{i13}g'^2f^2_Bm^2_{B^+}p_{cm}}{128\pi(m_u+m_b)^2M^4_{\tilde{f}}}
\left(M^2_{B^+}-m^2_{\ell_i}-M^2_{\cht^0_1}\right),
\label{eqn:B+}
\end{eqnarray}
where $p_{cm}$ is the momentum of the decay products in the rest frame
of the decaying meson, $m_{u,d,b}$ are the up, down and bottom quark
masses respectively, $m_{B^0}$ is the $B^0$ mass, $m_{B^+}$ is the
$B^+$ mass. Here $L_f = -g'Y_{f_L}/2$ for the left-handed fermions and
$R_f = g'Y_{f_R}/2$ for the right-handed fermions. $f_B$ is the
pseudo-scalar decay constant for $B$ decays, $M_{\cht^0_1}$ is the
lightest neutralino mass, $M_{\dnt_L}$ is the left down squark mass,
$M_{\upt_L}$ is the left up squark mass, and $M_{ \tilde{b}_R}$ is the
right bottom squark mass. In Eqns.\,(\ref{eqn:B0},\ref{eqn:B+}) we have
assumed that the sfermions have a common mass $M_{\tilde{f}}$.  The
pseudo-scalar decay constant for the B system has not been measured
experimentally and must be taken from lattice QCD. We have used the
value
\begin{equation} 
f_B =204\pm 30\ 
\rm{MeV}, 
\end{equation} 
from Ref.\cite{AliKhan:2001jg} where we have added the errors in
quadrature.  The branching ratio for the decay $B^0\ra\cht^0_1\bar{
  \nu}$ was taken as an input and the branching ratio for $B^+\ra\cht
^0_1\ell^+$ calculated from it using the above results. The same cuts
were applied as in the previous section.  The number of events which
would be observed in the detector is shown in Fig.\,\ref{fig:Bdecay}.
This shows that even for branching ratios below $10^{-7}$ there is a
significant range of neutralino lifetimes for which there are enough
events to explain the NuTeV results. The present experimental upper
limit on the branching ratio of the purely muonic decay is
$Br(B^\pm\ra\mu^\pm\nu_\mu)<2.1\cdot10^{-5}$ \cite{Groom:2000in}.

Using the results for the RPV branching ratios of the B mesons and the
neutralino lifetime we can find regions in $(\lam_{232},\lam'_{113})$
parameter space, for a given sfermion mass, in which there are $3\pm1$
events inside the NuTeV detector, this is shown in
Fig.\,\ref{fig:parameter}. We have included the low-energy bounds
Eq.(\ref{eq:bounds}).  In the case of the coupling $\lam'_{213}$ there
is also a bound on the product of the couplings $\lam_{232}\cdot\lam'
_{213}$ from the limits on the branching ratios for $B^0\ra\tau^-\mu
^+$ and $B^+\ra\mu^+\nu$ \cite{Groom:2000in}, the latter giving the
stricter bound
\begin{equation}
\frac{\lam'^2_{213}\lam^2_{232}f^2_Bm^5_{B^+}} {32\pi
  M^4_{\tilde{f}}\left(m_b+m_u\right)^2\Gamma_{B^+}}
\left(1-\frac{m_\mu^2}{m^2_{B^+}}\right)^2 \leq 2.1\times10^{-5}, 
\end{equation}
Here $\Gamma_{B^+}$ is the total width for the $B^+$. This gives
\begin{equation} 
\lam'_{213}\lam_{232} \leq3.8\times10^{-4}
\left(\frac{M_{\tilde{f}}}{200\,\rm{GeV}}\right)^2.  
\end{equation}

In Fig.~\ref{fig:parameter} we see that for every value of $\lam'_{i
  13}$ there are two solutions in $\lam_{232}$, except for a minimum
value of $\lam'_{i13}$, below which there are no solutions. This can
be understood as follows. The maximum fraction of neutralinos decays
in the distant detector for a lifetime $\tau=\beta c\gamma/\ell$, {\it
  i.e.} when the decay length corresponds to the flight length, the
distance between the production target and the detector. This
optimised lifetime corresponds numerically to
\begin{equation}
\lambda_{232} = 5.3\cdot10^{-4}
                  \left( \frac{M_{\tilde{f}}} {200\,\rm{GeV}} \right)^2. 
\end{equation}
This requires the minimum production rate and thus the minimum value
of $\lam'_{i13}$, which is the dip in the curves in
Fig.~\ref{fig:parameter}. For larger values of $\lam'_{i13}$ the
neutralino production is increased. We can then tune the lifetime of
the neutralino such that the decay length is either shorter or longer
than the flight length, yielding the two solutions shown in the
figure.

\subsection{$\tau$-Decays}

\begin{figure}
\begin{center}
\includegraphics[width=0.6\textwidth,angle=90]{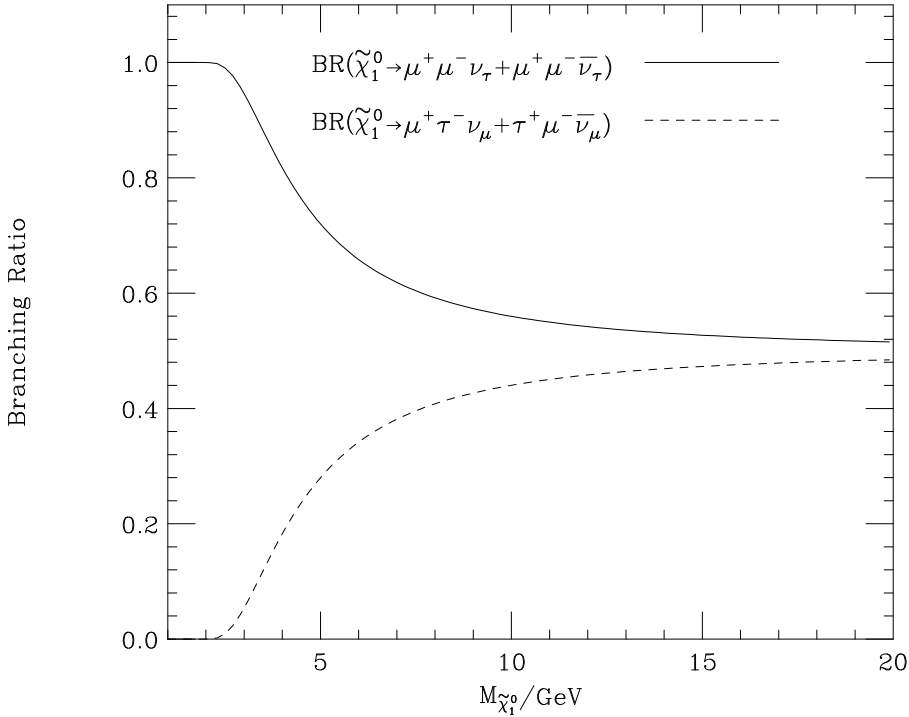}
\caption{Branching ratios for the decay of a purely bino lightest neutralino
        via the RPV coupling $\lam_{232}$. The sfermions have been assumed
         to be degenerate and light/right stau mixing has been neglected.}
\label{fig:tau}
\end{center}
\end{figure}

As discussed in section \ref{sec:model}, in our model the neutralino
can decay to $\mu\tau\nu$ as well as $\mu\mu\nu$. Using the
calculation of Eq.(\ref{eq:neutdecay}) we can compute the branching
ratios \linebreak\mbox{$Br_{\mu\mu} \equiv Br(\cht
  ^0_1\ra\mu^\pm\mu^\mp\nu_\tau)$} and \mbox{$Br_{\mu\tau}\equiv
  Br(\cht^0_1\ra \tau^\pm\mu^\mp \nu_\mu)$}, which are displayed in
Fig.\,\ref{fig:tau}. For neutralino masses above $10$-$15\gev$ the two
decays have practically equal branching ratios. However, when the
neutralino mass is close to the $\tau$-mass, $Br_{\mu\tau}$ is phase
space suppressed. For $M_{\chi^0_ 1}=5\,\rm{GeV}$ we have
$Br_{\mu\tau}=0.287$.  In obtaining Fig.\,\ref{fig:tau} the sfermions
have been assumed to be degenerate and left/right stau mixing has been
neglected.\footnote{In models where the scalar masses are unified at
  the GUT scale the running of the masses to low scales forces the
  right stau to be lighter than the left stau.  For low $\tan\beta$ it
  is a good approximation to neglect left/right stau mixing. For large
  values of $\tan\beta$ the right stau becomes much lighter, but this
  does not contribute to the decay.  It is thus a conservative
  assumption to require degenerate scalar fermion masses.} In
principle the NuTeV experiment can observe the $(\mu\tau\nu)$-modes
through the decays: $\tau^\pm\ra e^\pm\nu\nu$ and $\tau^\pm
\ra\pi^\pm(n\cdot \pi^0)\nu$, which would lead to unobserved $(e,\mu)$
and $(\pi,\mu)$ events, respectively.  Here $(n\cdot\pi^0)$ indicates
an additional $n=0,1,2,3$ emitted neutral pions.  Given the 3
observed $(\mu,\mu)$ events one would expect the following number of
events for $M_{\chi^0_1}=5\,\rm{GeV}$
\begin{eqnarray}
N_{(e,\mu)}&=&3\cdot\frac{Br_{\mu\tau}}{1-Br_{\mu\tau}}
\cdot Br(\tau\ra e\nu\nu)\approx 0.21 \\
N_{(\pi,\mu)}&=& 3\cdot\frac{Br_{\mu\tau}}
{1-Br_{\mu\tau}}\cdot Br(\tau\ra \pi
(n\cdot\pi^0)\nu)\approx 0.56,
\end{eqnarray}
where we have used the $\tau$ branching ratios from Ref.\cite{Groom:2000in}.
Thus the non-observation of $(e,\mu)$- and $(\pi,\mu)$-events is consistent.
We note that some of the $\tau\ra\pi^\pm(n\cdot\pi^0)\nu$ decays would show
extra activity in the detector and thus be rejected as pure $\pi^\pm$ events.
Therefore the above estimate is conservative \cite{conrad}.

\section{Future Tests of the $R_p$ Violating Model}
\subsection{NOMAD}

\begin{figure}
\begin{center}
\includegraphics[width=0.6\textwidth,angle=90]{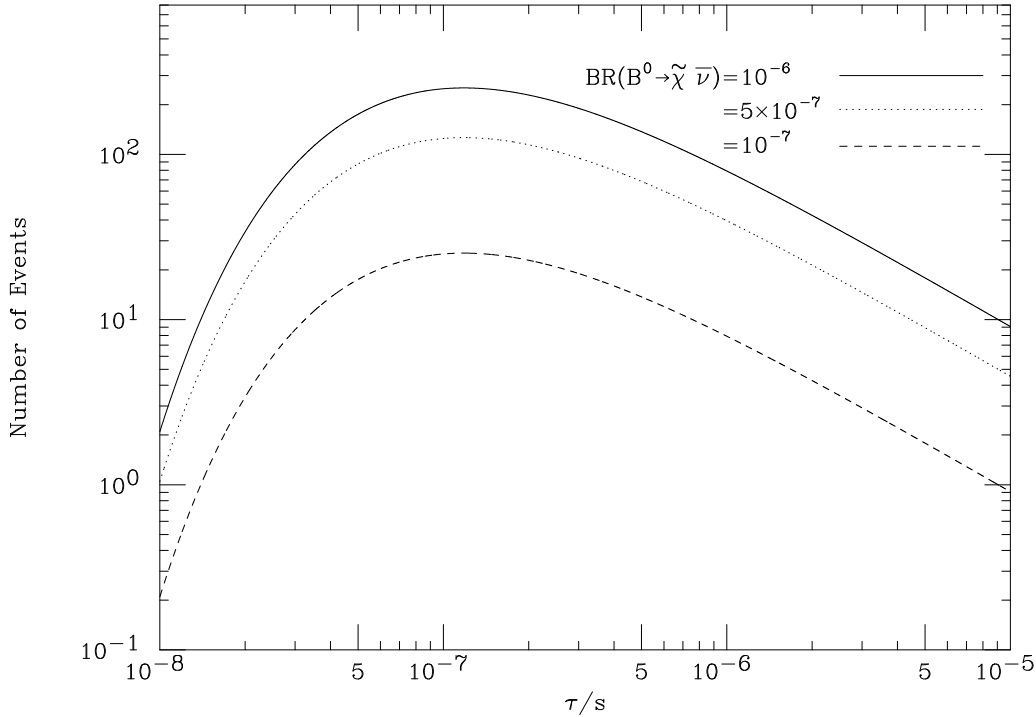}
\caption{The predicted number of di-muon events at NOMAD as a function of the
  neutralino lifetime. We have used our model for neutralino production
through $B$-meson decays. We indicate the prediction for three different
branching ratios of the neutral $B$-meson decay to neutralinos as in
Fig.\ref{fig:Bdecay}.}
\label{fig:nomad}
\end{center}
\end{figure}

The NOMAD experiment
\cite{Autiero:1997ui,Altegoer:1998gv,Astier:2001ck} was a neutrino
oscillation experiment at CERN which was dismantled in 1999.  The data
however are still on tape and could be used to test the current
proposal. We modified our program to estimate the event rate at NOMAD.
For this we used the following numbers
\cite{Autiero:1997ui,Altegoer:1998gv,Astier:2001ck}: distance
target-detector: $\ell=835\,$m, fiducial volume of the detector:
$V=(2.6\times2.6 \times 4)$ m$^3$, target material: Beryllium, target
density: $\rho=1.  85\,$ g/cm$^3$, target length: $d=1.1\,$m, proton
beam energy: $E=450\gev$, integrated number of protons: $N_p=4.1\cdot
10^ {19}$. Using these numbers we show our prediction for the number
of events at NOMAD in Fig.\ref{fig:nomad}.  For the same $B^0$-meson
branching ratio we obtain about an order of
magnitude more events than at NuTeV. Thus our model can be {\it
  completely} tested by the NOMAD data! 

The higher sensitivity at NOMAD can be understood as follows.  The
total $b{\bar b}$ production cross section for collisions on $Be$ at
NOMAD is $4.7\,$nb, while for collisions on $BeO$ at NuTeV it is
$94\,$nb. The total integrated luminosities are $5.58\cdot10^{11}
\,$nb$^{-1}$ (NOMAD) and $6.189\cdot10^{9}\,$nb$^{-1}$ (NuTeV),
respectively. Therefore the number of $b{\bar b}$-events are $2.6\cdot
10^{12}$ (NOMAD) and $5.8\cdot 10^{11}$ (NuTeV), respectively, {\it
i.e.}  about 4.5 times more at NOMAD. The NOMAD detector is closer and
than the NuTeV detector and thus subtends a larger solid angle by
about a factor of 3. The required neutralino lifetime is about the
same because NOMAD is about half the distance but the energy is also
about half. The NOMAD detector is about 8 times shorter but the
Lorentz boost is only about half the NuTeV boost, so this corresponds
to a factor of 4. All in all we would expect about a factor of 3.4
times more events at NOMAD than at NuTeV.  Comparing
Fig.~\ref{fig:nomad} with Fig.~\ref{fig:Bdecay} we see that this is
close to what the full numerical simulation gives.

\subsection{B-Factories}
As outlined above, for neutralino production we are relying on a rare
B-meson decay
\begin{eqnarray}
B^\pm&\ra&\mu^\pm+\chi^0_1,\label{eq:1dec}\\
B^0&\ra&\nu+\chi^0_1,\label{eq:2dec}
\end{eqnarray}
which can possibly be observed at a present or future B-factory. In
the Standard Model we have the decay $B^\pm\ra\mu^\pm+\nu$ with a
predicted branching ratio \cite{Harrison:1998yr} of about $3\cdot
10^{-7}$. This is probably just below visibility at BaBar
\cite{Harrison:1998yr}.

The decay (\ref{eq:1dec}) differs from the Standard Model decay
$B^\pm\ra\mu+\nu$ in the energy of the $\mu$, which is now only
$E_\mu=(M_{B^\pm}^2+m_\mu^ 2-M_\chi^2)/(2M_{B^\pm})\approx 0.27\gev$
for $M_\chi=5\gev$. We thus have a monochromatic muon with an order of
magnitude less energy than in the Standard Model decay. This is a
distinctive signature which we propose for investigation at BaBar and
other B-factories. We presume this is very difficult due to many
sources of soft muons as background. Also the efficiency for such soft
muons is typically very low, only about $5\%$ \cite{Harrison:1998yr}.

The decay (\ref{eq:2dec}) is invisible, with the neutralino decay far
outside the detector at a B-factory. If we have a $B^0$-${\bar B}^0$
system and could tag one of the mesons, via a conventional decay, then
we would have an unexpected invisible decay on the opposite side. We
propose this as a possible signature for investigation by the
experimental collaborations.

\section{Neutral Heavy Leptons}
\label{sec:nhl}

In \cite{Adams:2000rd,Adams:2001sk} the NuTeV collaboration also
considered the possibility of a neutral heavy lepton (NHL) to explain
their observation. Here a NHL, $N_{iL}, i=1,2,3$ is considered as a
primarily isosinglet field under $SU(2)_L$ with a small admixture of
the light Standard Model neutrinos. This is discussed for example in
Refs.~\cite{Gronau:1984ct,Johnson:1997cj}. We follow the notation of
Ref.~\cite{Gronau:1984ct}. In general such a NHL has charged current
(CC) and neutral current (NC) purely leptonic decays proceeding via a
virtual $W^\pm$ or $Z^0$-boson, respectively,
\begin{eqnarray}
N_{iL}&\ra&\ell_j^-+\ell_k^++{\bar\nu}_k,\quad (CC)\\
N_{iL}&\ra&\nu_m+\ell_n^++\ell_n^-.\quad (NC)
\end{eqnarray}
For the NC-decay the charged leptons are from the same family, whereas
for the CC-decay they can also be from different families. A given
CC-leptonic decay is proportional to the mixing element
$|U_{j{N_i}}|^2$. There is a corresponding NC-decay proportional to
the same mixing element for $m=j$. For a given set of NHL masses and
mixings, we typically would expect both NC- and CC-decays to occur.
$k,n=1,2,3$ are free indices which all contribute to the decay rate,
independent of the mixings.

NuTeV observe an excess of di-muon events. Assume we have one NHL,
$N_i$, with mass $M_{N_i}=5\gev$, and the other NHL's unobservably
heavy. The di-muon events could occur through CC-decays with $j=k=2$
and the mixing element $U_{2{N_i}}$ or through NC-decays with $n=2$
and the mixing elements $U_{m{N_i}}$, $m=1,2,3$. For $j=k =n=2$ we
obtain di-muon events through both NC- and CC-decays.

\begin{figure}[t]
\begin{center}
\includegraphics[angle=90,width=0.7\textwidth]{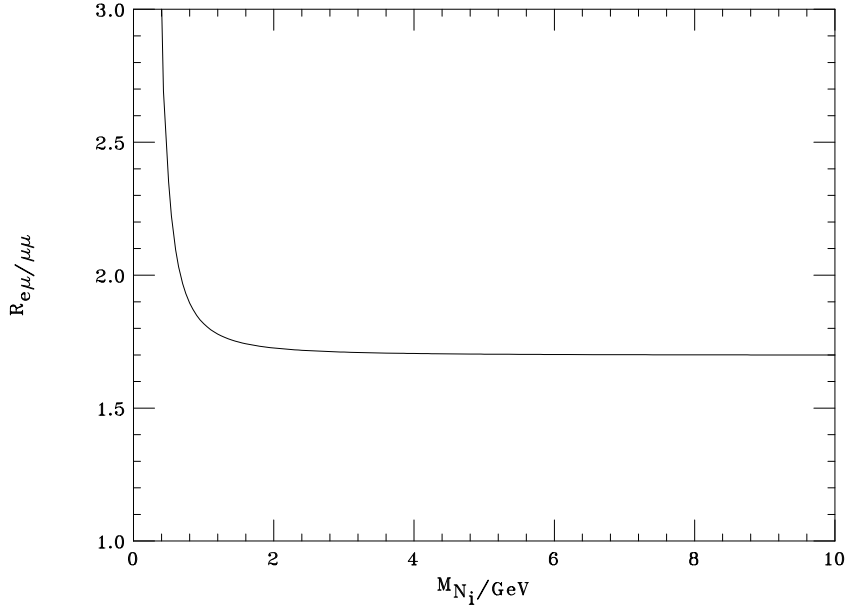}
\caption{Ratio of $(e,\mu)$- 
  to $(\mu,\mu)$-events, $R_{e\mu/\mu\mu}
  \equiv\frac{\Gamma(N_2\rightarrow
    e^+\mu^-\nu_\mu)}{\Gamma(N_2\rightarrow \mu^+\mu^-\nu_\mu)}$ in
  the decay of a NHL, $N_i$, versus the mass of $N_i$ .}
\label{fig:nhl}
\end{center}
\end{figure}

If the CC-decays contribute, {\it i.e.} $j=2$, we would expect there
to be accompanying $(e,\mu)$ events with similar probability, from the
$k=1$ mode. For example for a non-vanishing $|U_{2{N_i}}|^2$, using
the decay rates given in \cite{Johnson:1997cj}, we obtain the ratio of
$(e,\mu)$ to $(\mu,\mu)$ events given by $R_{e\mu/\mu\mu}\equiv
\Gamma(N_2\rightarrow e^+\mu^-\nu_\mu)/\Gamma(N_2\rightarrow
\mu^+\mu^-\nu_\mu)$. We plot this as a function of the NHL-mass in
Fig.\ref{fig:nhl}.
%%%%%%%%%%%%%%%%%%%%%%%%%%%%%%%%%%
%%%%%%%%%%%%%%%%%%%%%%%%%%%%%%%%%%
From the plot we see that we would expect {\it more} $(e\mu)$ events
than $(\mu,\mu)$-events.  This is excluded by the NuTeV non-observation 
of such events.

If the NC-decays contribute we can expect further $(e,e)$ and $(\tau
,\tau)$ events. The latter are kinematically suppressed as in the
$\not\!R_p$ case above. A search for the $(e,e)$-modes has only been
presented for low-energy electrons \cite{Formaggio:2000ne}.  However,
given a non-vanishing mixing element $U_{m{N_i}}$ which gives the
$(\mu,\mu)$-events via NC-decays we would expect further CC-decays:
$N_{iL}\ra\ell^\pm_j +\ell^\mp_k+\nu_k$, $k=1,2,3$. In particular, for
$j=1,2$ this leads again to $(e,\mu)$ events which were not observed.

We have thus eliminated all cases except a special model, which we
consider in more detail. Assume $j=3$ and $U_{3N_i}$ is the only
non-negligible mixing element. Furthermore, as above, assume $M_{N_
i}=5\gev$ and the other NHL's are unobservably heavy. We then have the
following decay modes
\begin{eqnarray}
N_i&\stackrel{ {\rm CC} }{\ra}& \left\{\tau\tau\nu_\tau,\tau\mu\nu_\mu,\tau e
\nu_e \right\},\\
N_i&\stackrel{ {\rm NC} }{\ra}& \left\{\nu_\tau ee,\nu_\tau\mu\mu,
\nu_\tau\tau\tau\right\}.
\end{eqnarray}
The $\tau$ and $\tau\tau$ decay modes are kinematically suppressed as
in the $\not\!\!R_p$ case discussed above and the observed di-muon
events are obtained form the NC-decay. This model has been studied by
the NOMAD collaboration for $M_{N_i}=10-190\,MeV$
\cite{Astier:2001ck}.

We now estimate the event rate for this model ($j=3$, $U_{3N_i}\not=0$). 
The production mechanism will go either via (CC or NC) Drell-Yan with
the tau neutrino mixing with the $N_i$ or via B-meson decays. We have
computed the Drell-Yan production cross section to be $\sigma_{DY}=
{\cal O} (10^{-1}\,{\rm pb})\cdot|U_{3N_i}|^2$. The neutral current
contribution to the NHL production is more than an order of magnitude
smaller. The total integrated luminosity at NuTeV corresponds to
$\sim6.2\cdot10^6\,{\rm pb}^{-1}$ giving the number of produced $N_i$:
$N_{N_i}^{prod}\sim4\cdot10^{6}\,|U_{ 3N_i}|^2$. Of these only about
$1\%$ fly in the direction of the detector \cite{Borissov:2000eu},
leaving us with $N_{N_i}\sim4\cdot10 ^{4}\,|U_{3N_i}|^2$. In order to
estimate the total number of events we must combine this with the
fraction of $N_i$ which decay in the detector given by
Eq.~(\ref{eq:decayprob}). The total event rate is proportional to
\begin{equation}
N_{ev}\approx N_{N_i} \exp\left\{-a|U_{3N_i}|^2\right\}\cdot b 
|U_{3N_i}|^2 =4\cdot 10^4 b\, |U_{3N_i}|^4 \exp\left\{-a|U_{3N_i}|^2\right\}
\end{equation}
where $a=\ell/(\beta\gamma c(\tau_{N_i}|U_{3N_i}|^2)$ and $b=\Delta
x/(\beta\gamma c \tau_{N_i}|U_{3N_i}|^2)$ from
Eq.~(\ref{eq:decayprob}) are independent of $|U_{3N_i}|$. The event
rate is maximal for $|U_{3N_i}|^2=2/a$. We obtain an upper limit
on the lifetime if we assume the NC decay is dominant. The latter we
determine through the scaled muon lifetime
\begin{equation}
\tau_{N_i}<\tau_\mu\,\left(\frac{m_\mu}{M_{N_i}}\right)^5 |U_{3N_i}|^{-2}=
9\cdot10^{-15}s\, |U_{3N_i}|^{-2}.
\end{equation} 
We then obtain $a=5.2\cdot10^8/\gamma$ and $b=1.3\cdot10^7/\gamma$.
For $\gamma=10$, for example, we obtain the maximal event rate for
$|U_{3N_i}|=9\cdot10^{-5}$, which is compatible with the independent
bound $\sum_i |U_{3N_i}|^2 < 0.016$ \cite{Bergmann:1999rg}. Following
Eq.~(\ref{eq:decayprob}) the total fraction decaying in the detector
is then roughly $1.1\%$.  Combining this with the previous estimate of
the number produced we get a total maximal number of events of about
$N_{ev}^{max}=5\cdot10^{-7}$, which is of course too small.

The reason this is so much smaller than in the supersymmetric model is
that the lifetime of the NHL is typically much shorter. Thus the NHL's
typically would decay well before the detector. We get the maximal
number of events when the lifetime is approximately the flight time.
For this we need a very small $|U_{3N_i}|$. Since we only have one
parameter in this model this feeds into the cross section resulting in
the very low rate. We do not expect the production via $B$-mesons to
help. The branching ratio is suppressed compared to the SM decay
branching ratio $Br(B^+\ra \tau^+ N_i)= |U_{3N_i}|^2 Br(B^+\ra
\tau^+\nu_\tau)\approx 7\cdot10^{-5}|U_{3N_i}|^2$ and thus also too
small.

\section{Conclusions}
We have reconsidered the NuTeV di-muon observation in the light of
supersymmetry with broken R-parity and neutral heavy leptons. We have
shown that it is not possible to obtain the observed event rate with
pair production of light neutralinos or via the production of neutral
heavy leptons. However, we have introduced a new production method of
neutralinos via $B$-mesons. Due to the copious production of
$B$-mesons in the fixed target collisions the observed di-muon event
rate can be easily obtained for allowed values of the R-parity
violating couplings.

The model we have proposed can be completely tested using current
NOMAD data. We suspect this is true of any model one might propose.
If the NOMAD search is negative our model is ruled out and the NuTeV
observation is most likely not due to physics beyond the Standard
Model.

It is worth pointing out that through this mechanism we have opened a
new sensitivity range in the R-parity violating couplings.  At
colliders we can probe the range where the neutralino decays in the
detector. For a photino neutralino this corresponds to
\cite{Dreiner:1997uz}
\begin{equation}
\lam>5\cdot 10^{-7}\sqrt{\gamma}\left(\frac{{\tilde m}}
{200\,{\rm GeV}}\right)^2\left(\frac{100\,{\rm GeV}}
{M_{\tilde\gamma}}\right)^{5/2}
=9\cdot 10^{-4}\sqrt{\gamma}\left(\frac{{\tilde m}}
{200\,{\rm GeV}}\right)^2\left(\frac{5\,{\rm GeV}}
{M_{\tilde\gamma}}\right)^{5/2}.
\end{equation}
Here we have substituted the light neutralino mass we are considering.
For significant boost factors we thus can probe couplings at most down
to $10^{-3}$. From Fig.\,\ref{fig:parameter} we see that for a 200 GeV
sfermion we can probe couplings down to about $5\cdot10^{-6}$, which
is more than two orders of magnitude smaller! It is thus worthwhile to
study the production of neutralinos via mesons at fixed target
experiments in more detail.

Before concluding we also note that one might worry that the lightest
supersymmetric Higgs boson would decay dominantly to the two light
neutralinos and thus be invisible. However, as with the $Z^0$ boson,
the Higgs does not couple to a Bino neutralino.

\vspace{0.4cm}

\noindent {\bf \large Acknowledgements}
\vspace{0.4cm}

{\it A.D. would like to acknowledge financial support from the Network
  RTN European Program HPRN-CT-2000-0014 ``Physics Across the Present
  Energy Frontier: Probing the Origin of Mass.'' and also the BSM team
  of Les Houches 2001 for discussions on the NuTeV observation. H.D.
  would like to thank G. Polesello for suggesting we consider the
  prediction for NOMAD. P.R.  would like to thank V.~Gibson, R.S.
  Thorne and B.R.~Webber for useful discussions and PPARC for
  financial support. We would like to thank L. Borissov, J. Conrad and
  M. Shaevitz for discussions on Ref.  \cite{Borissov:2000eu} and the
  NuTeV experiment.}

\end{document}